\documentclass[aip,cha,superscriptaddress]{revtex4-1}
\usepackage{hyperref}	
\usepackage{graphicx}
\usepackage{amsmath}
\usepackage{fullpage}
\usepackage{color}
\begin{document}

\title{The Dynamics of Hybrid Metabolic-Genetic Oscillators}
\author{Ed Reznik}
\affiliation{Department of Biomedical Engineering, Boston University, Boston MA 02215}
\author{Tasso J. Kaper}
\affiliation{Department of Mathematics and Statistics, Boston University, Boston MA 02215}
\author{Daniel Segr\`{e}}
\affiliation{Department of Biomedical Engineering, Boston University, Boston MA 02215}
\affiliation{Department of Biology, Boston University, Boston MA 02215}
\affiliation{Program in Bioinformatics, Boston University, Boston MA 02215}

\begin{abstract}
The synthetic construction of intracellular circuits is frequently hindered by a poor knowledge of appropriate kinetics and precise rate parameters. Here, we use generalized modeling (GM) to study the dynamical behavior of topological models of a family of hybrid metabolic-genetic circuits known as ``metabolators.'' Under mild assumptions on the kinetics, we use GM to analytically prove that all explicit kinetic models which are topologically analogous to one such circuit, the ``core metabolator,'' cannot undergo Hopf bifurcations. Then, we examine more detailed models of the metabolator. Inspired by the experimental observation of a Hopf bifurcation in a synthetically constructed circuit related to the core metabolator, we apply GM to identify the critical components of the synthetically constructed metabolator which must be reintroduced in order to recover the Hopf bifurcation. Next, we study the dynamics of a re-wired version of the core metabolator, dubbed the ``reverse'' metabolator, and show that it exhibits a substantially richer set of dynamical behaviors, including both local and global oscillations. Prompted by the observation of relaxation oscillations in the reverse metabolator, we study the role that a separation of genetic and metabolic time scales may play in its dynamics, and find that widely separated time scales promote stability in the circuit. Our results illustrate a generic pipeline for vetting the potential success of a potential circuit design, simply by studying the dynamics of the corresponding generalized model.
\end{abstract}

\maketitle

\begin{quotation}The engineering of biological circuits, synthetically constructed in the laboratory and designed to exhibit novel behaviors, has matured rapidly as a discipline over the past decade. A major challenge for synthetic biology is understanding how different cellular subsystems, composed of distinct components and operating at widely different speeds, interface and work together to generate coherent cellular behaviors. In particular, metabolic pathways can rapidly harvest energy and nutrients for reproduction, while genetic regulatory networks slowly control these pathways in response to environmental changes. Here, we study a generalized version of the metabolator, a unique synthetic circuit composed of both metabolic and regulatory components, previously constructed experimentally and shown to exhibit oscillations. In addition to exploring multiple alternative network topologies, our generalized approach overcomes the inherent uncertainty associated with precise kinetic details of biological circuits. Through this analysis we identify circuit components that are essential for producing sustained oscillations. Furthermore, we find that certain dynamical regimes of a rewired metabolator display unexpected oscillatory properties, such as the capacity to suddenly transition to high amplitude oscillations. This novel behavior could be tested experimentally, and lead to novel synthetic biology modules and applications.
\end{quotation}

\section{Introduction}
Synthetic biology is in need of theoretical tools to design circuits with a high probability of exhibiting specific, user-defined dynamical behaviors. Conventionally, efforts in this regard have focused on explicit kinetic formulations of the dynamics of a circuit. Indeed, deterministic models of the toggle switch \cite{Gardner2000} and the repressilator \cite{Elowitz2000}, two early examples of successful synthetic circuits, directly predicted the dynamical behavior of these constructs and likely influenced the final design of the circuits themselves. Perhaps the most prominent challenge in assembling these models was a fundamental lack of knowledge regarding kinetic details, including the mechanistic form of the rate laws, as well as the precise values of rate constants themselves.

The primary focus of this work is to develop a modeling framework for understanding the dynamics of a biological circuit for which we have little detailed knowledge of the true kinetics. In particular, we will try to derive generic results dependent exclusively on the ``topology'' of the circuit, rather than its detailed kinetics. To do so, we will focus our efforts on a particular system known as the metabolator, a synthetic circuit constructed in \textit{Escherichia coli K12} which exhibits limit cycle oscillations. At a coarse level of description, the metabolator consists of two metabolites which are interconverted between each other by two enzymes. One of the metabolites in the circuit acts as a regulator, repressing the expression of the enzyme catalyzing the metabolite's production and activating the expression of the enzyme catalyzing its degradation. In Ref. \cite{metab}, the authors observed that in conditions of sufficiently high metabolic inflow to the metabolator, a gene encoding one of the regulated enzymes exhibited sustained oscillations. The metabolator is particularly interesting because it remains (to our knowledge) the only example of a synthetic circuit containing a combination of both metabolic and genetic (transcriptional/translational) components. In all other cases, the building blocks of circuits have been entirely transcriptional and translational. The ``hybrid'' nature of the metabolator is particularly interesting because metabolism is well-known to take place at a faster rate than transcription and translation. The role such separation of time scales actually plays in the dynamics of the metabolator is unresolved, although it has been observed to be quite in important in other metabolic contexts \cite{Jamshidi2008}.

One of our primary approaches to studying the metabolator is based on a non-dimensionalization technique for studying dynamical systems commonly referred to as generalized modeling (GM) \cite{SKM, GM, SKM2, SKM3,Gehrmann2011}. In GM, a change of variables is applied to a dynamical system so that many of the otherwise unconstrained parameters in the system (\textit{e.g.}, rate constants) are replaced by ``elasticity'' parameters with well-defined ranges. Then, the dynamics of the system can be studied around an arbitrary and potentially unknown steady-state. GM's conclusions are invariant to the particular choice of rate law, and depend only on the sensitivity of the rate law at the system's equilibrum. This enables the study of a \textit{topological} model of the metabolator, in which no assumptions are generally made on the explicit form of the rate law, nor the true values of the parameters in the system. In practice, many GM studies have focused on using large-scale numerical simulations to tease out statistical properties relating elasticities to stability properties of the system. In this work, we supplement such numerical work with complementary analytical studies, and highlight the additional insight that can be gained through explicit calculations. We complement our use of GM with explicit dynamical modeling of the metabolator, and highlight the complementary conclusions that can be drawn from the two techniques (e.g. local vs. global bifurcation phenomena when applying GM or explicit dynamical modeling, respectively). Importantly, we demonstrate how the results from GM can prompt further study with explicit modeling, and vice versa.

The main results of the work are as follows. First, a simple, ``core'' model of the metabolator (two genes, two metabolites) originally illustrated in Ref. \cite{metab} is investigated using a conventional dynamical model. We find that the system lacks the capability to pass through a Hopf bifurcation. Second, using GM, we analytically prove that \textit{any} dynamical model with a topology analogous to the core metabolator is incapable of Hopf bifurcations. Concluding that additional topological elements are necessary for a Hopf bifurcation, we use numerical simulations with GM to progressively restore elements of the metabolator model from Ref. \cite{metab} which we omitted in our core model. Doing so, we reveal the crucial topological components of the metabolator which endow it with the ability to oscillate. Third, we investigate the possibility of oscillations in an explicit model of a simply ``re-wired'' version of the metabolator in which the regulatory connections are reversed (the reverse metabolator, ``RM''). We show that the RM is capable of sustained oscillations by both a local Hopf bifurcation as well as global fold of limit cycles. Fourth, prompted by findings in the explicit model and using the GM tools we developed earlier, we study the role that metabolic and genetic time scales play in generating instability and potential oscillations in the RM \textit{in vivo}. 

\begin{figure}[ht]
	\centering
		\includegraphics[width=0.7\columnwidth]{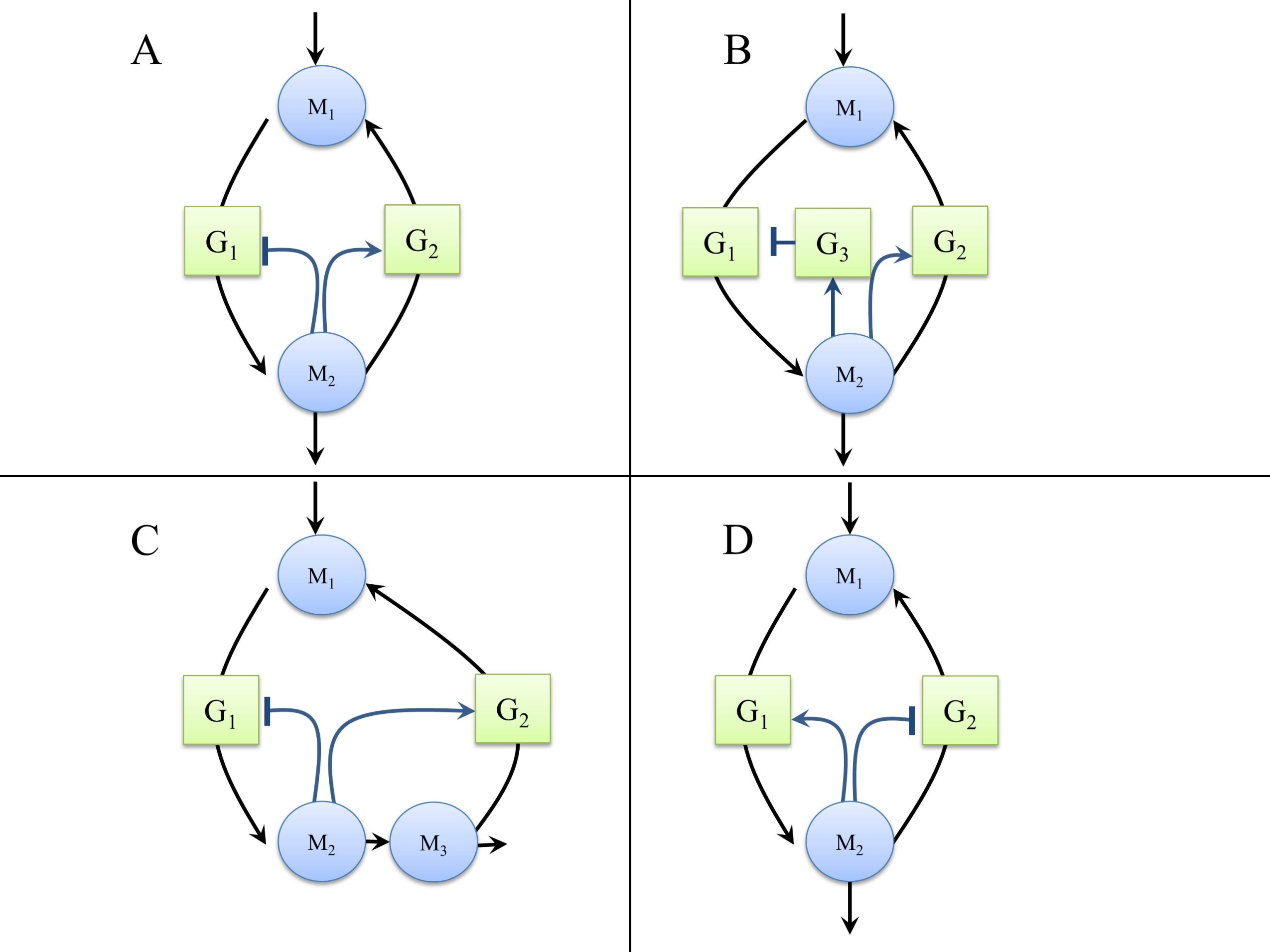}
	\label{fig:metabolator}
	\caption{Several different metabolator designs. (a) The core metabolator, studied in Sections \ref{sec:core}-\ref{sec:SKM}. Metabolite $M_2$ represses the transcription of gene $G_1$ and activates the transcription of gene $G_2$. (b) Core metabolator with intermediary delay gene $G_3$, and (c) Core metabolator with un-lumped metabolites, both studied in Section \ref{sec:alt}. (d) Reverse metabolator, studied in Section \ref{sec:RM}. In contrast to the core metabolator, $M_2$ activates the transcription of $G_1$ and represses the transcription of $G_2$.}
\end{figure}

\section{The Core Metabolator}
\label{sec:core}
To begin to understand the fundamental components which generate sustained oscillations in the metabolator, we developed a four-dimensional dynamical model of the system. Our particular choice of design was motivated by the schematic provided in Ref. \cite{metab} and re-illustrated in \ref{fig:metabolator} A. Two enzymes ($G_1$ and $G_2$) interconvert two substrates ($M_1$ and $M_2$). Gene $G_1$ codes for the enzyme which converts $M_1$ into $M_2$, and gene $G_2$ codes for the enzyme which converts $M_2$ into $M_1$. We explicitly assume that the rate of gene expression is substantially faster than protein translation, so that the abundances of the mRNA transcripts encoding $G_1$ and $G_2$ may be assumed to be in quasi-steady-state. Metabolite $M_2$ represses the expression of gene $G_1$ and activates the expression of gene $G_2$. Furthermore, we assume that the rate of inflow of $M_1$ is constant, and the rate of outflow of $M_2$ is in proportion to its concentration. Because of the symmetry in the dynamics of $M_1$ and $M_2$ (both are produced and degraded by the same metabolic reactions), we replace $M_1$ with the total amount of substrate in the system $C = M_1 + M_2$. Furthermore, we assume bilinear mass-action rate laws for the dynamics of the metabolic reactions. We dub our simplified model the \textit{core metabolator}. The nondimensional equations describing the dynamics of the core metabolator are

\begin{subequations}
	\begin{equation}
	\frac{dg_1}{dt} = \frac{1}{1+m_2^2} - g_1 
	\label{eq:met1_g1}
	\end{equation}
	\begin{equation}
	\frac{dg_2}{dt} = \frac{m_2^2}{1+m_2^2} - g_2 
	\label{eq:met1_g2}
	\end{equation}
	\begin{equation}
	\frac{dc}{dt} = \rho(1-m_2) 
	\label{eq:met1_c}
	\end{equation}
	\begin{equation}
	\frac{dm_2}{dt} = -\rho m_2 + \eta_1g_1(c-m_2) - \eta_2 g_2m_2,
	\label{eq:met1_n}
	\end{equation}
\label{eq:met_linear}
\end{subequations}

\noindent where the parameters $\rho, \eta_1, \eta_2$ are assumed to be strictly positive. See Appendix A for a derivation of (\ref{eq:met_linear}). We emphasize that the core metabolator serves as a jumping off-point: observing oscillations here would suggest the presence of oscillations in more detailed models capturing the true topological elements of the \textit{in vivo} metabolator. On the other hand, not observing oscillations in this simplified model implies that one needs to examine a model with more components in order to observe Hopf bifurcations, see Section 4. 

The fixed point of (\ref{eq:met_linear}) is $g_1^* = g_2^* = \frac{1}{2}, m_2^* = 1$ and $c^* = \frac{\eta_1+\eta_2+2\rho}{\eta_1}$. The Jacobian $J$, evaluated at the fixed point, becomes \\

\begin{equation}
\left[ \begin{array}{cccc}
-1 & 0 & 0 & \frac{-1}{2} \\
0 & -1 & 0 & \frac{1}{2} \\
0 & 0 & 0 & -\rho \\
2\rho+\eta_2 & -\eta_2 & \frac{\eta_1}{2} & \frac{-2\rho-\eta_1-\eta_2}{2} \end{array} \right]
\label{eq:J_coredim}
\end{equation}
The characteristic equation for $J$ is
\begin{equation}\label{eq:char_met1}
(1+\lambda)\left( \lambda^3+\lambda^2(\frac{2\rho+\eta_1+\eta_2+2}{2}) + \lambda(\frac{4\rho+3\eta_2+\eta_1+\eta_1\rho}{2}) +\frac{\eta_1\rho}{2} \right) =0.
\end{equation}
In order to identify potential candidates for a Hopf bifurcation, we make use of a constraint relating the coefficients of (\ref{eq:char_met1}) which must be satisfied in order for (\ref{eq:char_met1}) to have a pair of purely imaginary roots. First, we note that (\ref{eq:char_met1}) has a root at $\lambda = -1$. Then, we can substitute $\lambda_{1,2} = \pm i\omega$ into the cubic polynomial in (\ref{eq:char_met1}) we find that these conditions are $c_0 = \omega^2c_2 \text{ and } c_1 = \omega^2 c_3$, where $c_i$ is the coefficient of the $\lambda^i$th term of the cubic polynomial in parentheses in (\ref{eq:char_met1}). After substituting the appropriate coefficients into these conditions and some algebra, we find 

\begin{equation}
\frac{\eta_1\rho}{2+\eta_1+\eta_2+2\rho} = \frac{\eta_1+3\eta_2+4\rho+\eta_1\rho}{2}.
\label{eq:condition_met1}
\end{equation}
It is not clear if (\ref{eq:condition_met1}) can be satisfied with real, positive values of the system parameters. Rewriting (\ref{eq:condition_met1}) as a quadratic polynomial in $\eta_1$, we obtain
\begin{equation}
(1+\rho)\eta_1^2 + \left((1+\rho)(2+2\rho+\eta_2) + 2\rho + 3\eta_2 \right)\eta_1 + (2+2\rho+\eta_2)(4\rho + 3\eta_2) = 0.
\label{eq:quad_et1}
\end{equation}
Now the outcome becomes plainly clear: because (\ref{eq:quad_et1}) only contains positive coefficients, Descartes' Rule of Signs immediately proves that it cannot have positive real roots. Thus, there exists no combination of $(\eta_1,\eta_2,\rho)$ which will yield a pair of purely imaginary eigenvalues, and the system does not exhibit a Hopf bifurcation.

\section{Generalized Modeling of the Core Metabolator}
\label{sec:SKM}
Can we say anything more general about the oscillatory capabilities of the core metabolator illustrated in Figure 1A? In the previous section, we showed that one particular realization of this schematic model did not exhibit a Hopf bifurcation, but it remains unclear whether this was a quality of our choice of model. One may naturally wonder if choosing different kinetic rate laws for the metabolic reactions, or different regulatory kinetics for the feedback regulation, would yield qualitatively different behavior like oscillations. In this section, we propose a method for addressing these concerns by applying a technique known as generalized modeling (GM).

GM is a nondimensionalization procedure for reformulating the kinetics of a dynamical system, enabling the study of local dynamics near an equilibrium in an efficient way. To do so, parameters in the system are rewritten in terms of normalized parameters known as \textit{elasticities}. The elasticities have a direct connection to the original kinetic parameters, but are much easier to work with. Most importantly, elasticities typically have well-defined and limited ranges (e.g. [0,1]), and sampling them across this range effectively samples all possible values of the original kinetic parameters (which may otherwise span several potentially unknown orders of magnitude). Furthermore, these elasticities are not necessarily tied to any single choice of kinetic rate law; instead, they simply represent the sensitivity of any rate law to small changes in the equilibrium value of a variable, normalized by the steady-state magnitude of that rate law. In the context of metabolism, elasticities are similar to logarithmic derivatives of the rate law with respect to a change in substrate concentration \cite{Fell1992, cycle}.

This section is organized as follows. First, we introduce a motivating example of the GM nondimensionalization procedure. Next, we illustrate the role this non-dimensionalization plays in the formulation of generalized models. Then, we apply this procedure to generate a GM of the core metabolator. Finally, we analyze the stability properties of the GM, and analytically prove that a topological model of the core metabolator with fairly general rate laws is incapable of a Hopf bifurcation. 

\subsection{Generalized Modeling}
First, we present the normalization procedure upon which GM is based, known as generalized modeling, see for example Ref. \cite{GM}. Consider a system of ordinary differential equations,

\begin{align}
\frac{dX}{dt} & = I(X,Y) - O(X,Y) \nonumber \\
\frac{dY}{dt} & = F(X,Y) - G(X,Y)
\label{eq:motivating_1}
\end{align}
Here, $X$ and $Y$ are both functions of time. The stability of (\ref{eq:motivating_1}) around its equilibrium is determined by the eigenvalues of the Jacobian $J$. Here, we approach the problem of calculating $J$ after first executing a particular change of variables central to GM. Assume that there exists an equilibrium $(X_0,Y_0)$ and normalize (\ref{eq:motivating_1}) by its steady state concentration (assuming $I(X_0,Y_0), O(X_0,Y_0) \neq 0$):

\begin{equation}
\frac{1}{X_0}\frac{dX}{dt} = \frac{1}{X_0} \left( \frac{I(X_0,Y_0)}{I(X_0,Y_0)}I(X,Y) - \frac{O(X_0,Y_0)}{O(X_0,Y_0)}O(X,Y) \right).
\label{eq:motivating_2}
\end{equation}
Letting $x = \frac{X}{X_0}, y = \frac{Y}{Y_0}$ so that the variables are nondimensionalized and the equilibrium is at $(1,1)$, we can write 

\begin{equation}
\frac{dx}{dt} = \frac{1}{X_0} \left( \frac{I(X_0,Y_0)}{I(X_0,Y_0)}I(xX_0,yY_0) - \frac{O(X_0,Y_0)}{O(X_0,Y_0)}O(xX_0,yY_0) \right).
\label{eq:motivating_4}
\end{equation}

To find the components of the Jacobian for $\frac{dx}{dt}$, we simply differentiate each term by each independent variable and evaluate at the equilibrium $(x,y) = (1,1)$. If we let $h = \frac{dx}{dt}$ and write the components of the Jacobian as $\frac{\partial h}{\partial x}|_{(1,1)}$ and $\frac{\partial h}{\partial y}|_{(1,1)}$, we find 

\begin{align}
\frac{\partial h}{\partial x}|_{(1,1)} &= \frac{1}{X_0} \left( I(X_0,Y_0)\theta^I_x - O(X_0,Y_0)\theta^O_x \right) \nonumber \\
\frac{\partial h}{\partial y}|_{(1,1)} &= \frac{1}{X_0} \left( I(X_0,Y_0)\theta^I_y - O(X_0,Y_0)\theta^O_y \right),
\label{eq:motivating_5}
\end{align}
where $\theta^I_x = \frac{1}{I(X_0,Y_0)} \left( \frac{\partial I(xX_0,yY_0)}{\partial x} \right)$ and $\theta^O_x = \frac{1}{O(X_0,Y_0)} \left( \frac{\partial O(xX_0,yY_0)}{\partial x} \right)$. By repeating the calculations outlined above for $\frac{dy}{dt}$, we can calculate the remaining entries of the Jacobian. 

We argue that these $\theta$'s (herein referred to as \textit{elasticities} and similar to the elasticities from metabolic control analysis \cite{Fell1992}) have well-defined and limited ranges which makes performing calculations with them much easier than with similar calculations with the original parameters. To demonstrate the utility of elasticities, we will reformulate the dynamics of $G_1$ in the core metabolator. To do so, we will begin with the first equation of the dimensionalized dynamics of the metabolator, reproduced from Appendix A (see (\ref{eq:dim_g1})),

\begin{equation}
	\frac{dG_1}{dt} = \frac{\alpha_1}{K_1^2 + M_2^2} - \beta_1G_1.
	\label{eq:GM_g1}
\end{equation}

We let $g_1 = \frac{G_1}{G_{1,0}}$ and $m_2 = \frac{M_2}{M_{2,0}}$ (where $G_{1,0}$ and $M_{2,0}$ are the steady-state concentrations of $G_1$ and $M_2$, respectively) . Then, we can write
\begin{equation}
\frac{dg_1}{dt} = \frac{1}{G_{1,0}} \left(\frac{\alpha_1}{K_1^2 + (m_2M_{2,0})^2} - \beta_1g_1G_{1,0} \right).
\label{eq:elast1}
\end{equation}
We will calculate the elasticity of the first term in (\ref{eq:elast1}) corresponding to the production of $g_1$. To do so, we normalize this production term by its magnitude at steady state:
\begin{align*}
\mu(m_2) = \frac{ \frac {\alpha_1} {K_1^2+m_2^2M_{2,0}^2}} { \frac {\alpha_1} {K_1^2+M_{2,0}^2}} = \frac {K_1^2 + M_{2,0}^2} {K_1^2 + m_2^2M_{2,0}^2}.
\end{align*}

To calculate the elasticity, we take the derivative of $\mu(m_2)$ with respect to $m_2$, evaluated at $m_2=1$,
\begin{align*}
\frac{d\mu}{dm_2} \Big |_{m_2=1}&= \frac{-(K_1^2 + M_{2,0}^2)(2m_2M_{2,0}^2)}{\left(K_1^2 + m_2^2M_{2,0}^2\right)^2} \Big |_{m_2=1} \\
&=\frac{(-2M_{2,0}^2)(K_1^2+M_{2,0}^2)}{(K_1^2+M_{2,0}^2)^2}\\
& = \frac{-2M_{2,0}^2}{K_1^2 + M_{2,0}^2}.
\end{align*}
Upon inspection, this elasticity can only take values in $[-2,0]$. When $M_{2,0} \gg K_1$, the elasticity approaches negative two and the production of $G_1$ becomes very sensitive to small changes in the abundance $M_{2,0}$. In contrast, when $M_{2,0} \ll K_1$, the elasticity approaches zero and the production of $G_1$ is insensitive to changes in $M_{2,0}$. 

The well-defined and limited range of elasticities becomes extremely powerful in studying the dynamics of a system near its equilibrium. As we show in the ensuing section, these elasticities feature prominently in the Jacobian of a dynamical system reformulated with GM. We will demonstrate how elasticities can be used to prove the generic stability of a \textit{topological} model of a dynamical system, with only mild assumptions on the explicit rate laws governing the dynamics. 

\subsection{Generalized Model of the Core Metabolator}
We can use the aforementioned normalization procedure to develop a generalized model of the core metabolator. As we will show, the generalized model enables us to draw conclusions about a topological model of the core metabolator, while relaxing many of the assumptions we made in our explicit model of the core metabolator in Section 2. We begin with the following ``generalized'' form of the dynamics:

\begin{align}
	\frac{dM_1}{dt} &= I + G_2R_2(M_2) - G_1R_1(M_1) \nonumber \\
	\frac{dM_2}{dt} &= G_1R_1(M_1) - G_2R_2(M_2) - R_3(M_2) \nonumber \\
	\frac{dG_1}{dt} &= P_1(-M_2) - D_1(G_1) \nonumber \\
	\frac{dG_2}{dt} &= P_2(M_2) - D_2(G_2).
\label{eq:SKM_core1}
\end{align}

\noindent The terms $R_1$ and $R_2$ correspond to the reactions catalyzed by $G_1$ and $G_2$, respectively, while $I$ and $R_3$ correspond to the inflow and outflow of mass from the metabolator. The $P_i$ and $D_i$ terms correspond to the production and degradation of proteins. Importantly, the $P_i$ terms capture the feedback regulation of metabolite $M_2$ on the two enzymes. 

We make a few assumptions regarding the dynamics of the core metabolator in our generalized model. First, we assume $R_i, P_i$ and $D_i$ are generic, monotonically increasing functions of their arguments (i.e. $\frac{\partial R_1}{\partial M_1} > 0$). Many frequently used kinetic rate laws, including Hill, Michaelis-Menten, and mass-action kinetics satisfy this assumption, among many other possible forms of kinetics. We also retain the earlier assumption that the rates of metabolic reactions are linear with respect to the amount of gene product. The structure of (\ref{eq:SKM_core1}) corresponds precisely to the topological structure of the core metabolator in Figure 1A. 

Before we proceed to calculating the Jacobian of (\ref{eq:SKM_core1}), we briefly describe a simple method for calculating the prefactors (those parameters which were not elasticities, i.e. $O(X_0,Y_0)$) which appear in the Jacobian and were described in the prior section. In prior work on GM \cite{SKM}, it has been shown that for metabolic networks, these prefactors correspond precisely to the rates of metabolic reactions at equilibrium. As we show below, these rates can be calculated by enforcing that at equilibrium the fluxes of reactions flowing into and out of each metabolite are balanced \cite{Orth2010}. Furthermore, for the non-metabolic components of the metabolator, we can apply a similar balancing procedure by noting that at equilibrium, the production and degradation rates for each protein must be equal. Applying these balancing conditions to (\ref{eq:SKM_core1}), we find that 

\begin{align}
G_{1,0}R_1(M_{1,0}) - G_{2,0}R_2(M_{2,0}) &= I \nonumber \\
G_{1,0}R_1(M_{1,0}) - G_{2,0}R_2(M_{2,0}) &= R_3(M_{2,0}) \nonumber \\
P_1(-M_{2,0}) &= D_1(G_{1,0}) \nonumber \\
P_2(M_{2,0}) &= D_2(G_{2,0}).
\label{eq:constraints}
\end{align}

Here, $M_{1,0}, M_{2,0}, G_{1,0}$ and $G_{2,0}$ indicate the steady state concentrations of metabolites 1 and 2, and gene products 1 and 2, respectively. The first two constraints in (\ref{eq:constraints}) force the rate of reaction $G_1R_1$ (corresponding to the conversion of $M_1$ to $M_2$ ) to be precisely balanced by the rate of reaction $G_2R_2$ (the conversion of $M_2$ to $M_1$) and $I$, the rate of inflow of $M_1$ to the system. Similarly, $G_2R_2$ must be precisely equal to the sum of $G_1R_1$ and $R_3$ (the outflow of $M_2$ from the system). The final two conditions impose that, for each gene, the rates of production and degradation must be balanced.

Because of the interdependency of all the prefactors in (\ref{eq:constraints}), it becomes useful to explicitly introduce the parameter $v=G_{2,0}R_2(M_{2,0})$. We also introduce a free parameter $\alpha>0$ by setting $\alpha v = R_3(M_{2,0})$. From the equilibrium condition (\ref{eq:constraints}), it follows that $I=\alpha v \text{ and } G_{1,0}R_1(M_{1,0}) = (1+\alpha)v$. Finally, we let $L_1 = P_1(-M_{2,0}) = D_1(G_{1,0}), L_2 = P_2(M_{2,0}) = D_2(G_{2,0})$.  We can now rescale (\ref{eq:SKM_core1}) in the same manner as (\ref{eq:motivating_4}):
\begin{align}
	\frac{dm_1}{dt} &= 	\frac{1}{M_{1,0}} \left(\alpha v + \frac{v}{G_{2,0}R_2(M_{2,0})}g_2G_{2,0}R_2(m_2M_{2,0}) - \frac{(1+\alpha)v}{G_{1,0}R_1(M_{1,0})}g_1G_{1,0}R_1(m_1M_{1,0}) \right) \nonumber \\
	\frac{dm_2}{dt} &= 	\frac{1}{M_{2,0}} \left( \frac{(1+\alpha)v}{G_{1,0}R_1(M_{1,0})}g_1G_{1,0}R_1(m_1M_{1,0}) - \frac{v}{G_{2,0}R_2(M_{2,0})}g_2G_{2,0}R_2(m_2M_{2,0}) \right. \nonumber \\
& \left. -  \frac{\alpha v}{R_3(M_{2,0})}R_3(m_2M_{2,0}) \right) \nonumber \\
	\frac{dg_1}{dt} &= \frac{1}{G_{1,0}}\left( \frac{L_1}{P_1(-M_{2,0})} P_1(-m_2M_{2,0}) - \frac{L_1}{D_1(G_{1,0})}D_1(g_1G_{1,0}) \right) \nonumber \\
	\frac{dg_2}{dt} &= \frac{1}{G_{2,0}}\left(  \frac{L_2}{P_2(M_{2,0})}P_2(m_2M_{2,0}) - \frac{L_2}{D_2(G_{2,0})}D_2(g_2G_{2,0})	\right),
\label{eq:SKM_core2}
\end{align}
where $m_1 = \frac{M_1}{M_{1,0}}, m_2 = \frac{M_2}{M_{2,0}}, g_1 = \frac{G_1}{G_{1,0}}, g_2 = \frac{G_2}{G_{2,0}}$.

The Jacobian matrix $J$ of the core metabolator is then

\begin{equation}
J = \left[ \begin{array}{cccc}
\frac{-\theta_1(1+\alpha)v}{M_{1,0}} & \frac{\theta_2v}{M_{1,0}} & \frac{-(1+\alpha)v}{M_{1,0}} & \frac{v}{M_{1,0}} \\
\frac{\theta_1(1+\alpha)v}{M_{2,0}} & \frac{-\theta_2v-\alpha\theta_3v}{M_{2,0}} & \frac{(1+\alpha)v}{M_{2,0}} & \frac{-v}{M_{2,0}} \\
0 & \frac{L_1\theta_{f1}}{G_{1,0}} & -\frac{L_1\theta_{d1}}{G_{1,0}} & 0 \\
0 & \frac{L_2\theta_{f2}}{G_{2,0}} & 0 & -\frac{L_2\theta_{d2}}{G_{2,0}} \end{array} \right].
\label{eq:J_core}
\end{equation} 

The parameters $\theta_i$ in (\ref{eq:J_core}) correspond to the elasticities of metabolic reactions described earlier: 
\begin{align}
\theta_1 &= \frac{1}{G_{1,0}R_1(M_{1,0})} \left( \frac{\partial ( g_1G_{1,0} R_1( m_1M_{1,0}) ) }{\partial m_1} \right) \Big|_{m_1 = g_1 = 1} \nonumber \\
\theta_2 &= \frac{1}{G_{2,0}R_2(M_{2,0})} \left( \frac{\partial ( g_2G_{2,0} R_2( m_2M_{2,0} ) )}{\partial m_2} \right) \Big|_{m_2 = g_2 = 1} \nonumber \\
\theta_3 &= \frac{1}{R_3(M_{2,0})} \left( \frac{ \partial R_3(m_2M_{2,0}) }{\partial m_2} \right) \Big|_{m_2 = 1} .
\label{eq:theta1}
\end{align}

\noindent Here, $\theta_i>0, i =1,2,3 $ follows directly from the earlier assumption that all $P_i,D_i, R_i$ are monotonically increasing functions of their arguments. Similarly, the parameters $\theta_{f1}, \theta_{f2}, \theta_{d1}, \theta_{d2}$  in (\ref{eq:J_core}), corresponding to elasticities of the regulatory reactions, can be calculated by:
\begin{align}
\theta_{f1} &= \frac{1}{P_1(-M_{2,0})} \left( \frac{\partial P_1(-m_2M_{2,0})}{\partial m_2} \right) \Big|_{m_2 =1} \nonumber \\
\theta_{f2} &= \frac{1}{P_2(M_{2,0})} \left( \frac{\partial P_2(m_2M_{2,0})}{\partial m_2} \right) \Big|_{m_2 =1} \nonumber \\
\theta_{d1} &= \frac{1}{D_1(G_{1,0})} \left( \frac{\partial D_1(G_{1,0})}{\partial g_1} \right) \Big|_{g_1 =1} \nonumber \\
\theta_{d2} &= \frac{1}{D_2(G_{2,0})} \left( \frac{\partial D_2(G_{2,0})}{\partial g_2} \right) \Big|_{g_2 =1}.
\label{eq:thetar1}
\end{align}

\noindent For the generalized model of the core metabolator , $\theta_{f1}<0, \theta_{f2}>0$ since $M_2$ inhibits $G_1$ and activates $G_2$. Furthermore, $\theta_{d1},\theta_{d2}>0$ since an increase in the abundance of a gene product leads to an increase in its rate of degradation.

The utility of reformulating the core metabolator (\ref{eq:met_linear}) with the GM in (\ref{eq:SKM_core2}) is now plainly clear. Many of the prior assumptions made on the rate laws in our explicit dynamical model (\ref{eq:met_linear}) have been relaxed. Our sole assumption is that the rate laws are monotonic, so that it is always the case that the signs of the elasticities are known.  Now, by analytically studying the eigenvalues of $J$, we can understand the stability properties of the core metabolator across the entire spectrum of possible parameter values and rate laws. 

In Appendix B, we show that (\ref{eq:J_core}) cannot have eigenvalues with positive real part for any admissible value of parameters. The proof in the Appendix follows by an application of the Routh-Hurwitz criterion, a well-known tool from classical control theory \cite{Astrom2008}. Since (\ref{eq:J_core}) does not have any eigenvalues with positive real part, we have proven that all equilibria of the GM of the core metabolator are exponentially stable to small perturbations. This directly implies that the GM of the core metabolator cannot undergo a Hopf bifurcation.

The main result of this section bears some elaboration: the GM of the core metabolator encompasses all possible explicit realizations of the topological model shown in Figure \ref{fig:metabolator}A, under the general hypotheses about the rate laws stated above. With elementary techniques, it was possible to show that none of these potential realizations could ever pass through a Hopf bifurcation. We note that there may still be global mechanisms, such as a saddle node of limit cycles, that introduce oscillations. This naturally suggests that some additional component of the topology of the \textit{in vivo} metabolator, which has been removed in our model of the core metabolator, must be crucial to the existence of a Hopf bifurcation. This idea is explored further in the following section.

\section{Hopf Bifurcations in More Complete Generalized Models of the Metabolator}
\label{sec:alt}
The analysis in Sections 2 and 3 is suggestive: which topological simplifications in the core metabolator and in the GM of the core metabolator eliminated the possibility of a Hopf bifurcation? To address this question, we used GM to inspect which components would lead to oscillations when reintroduced into the core metabolator. To do so, successively more complete versions of the \textit{in vivo} metabolator were created, as shown in Figure 1B-C. We developed two hypotheses for the crucial topological features of the metabolator which would be necessary for sustained oscillations. First, we reasoned that a third gene, labeled $G_3$ and representing LacI from Ref. \cite{metab}, might act as an intermediary delay in the negative feedback from $M_2$ to $G_1$ and consequently induce oscillations (see Figure 1B). Such delays in feedback have been implicated in the literature as a major source for the onset of oscillations. Second, we separated the lumped metabolite $M_2$ into its appropriate constitutive molecules (now labeled $M_2$ and $M_3$) in Figure \ref{fig:metabolator}C). In the revised topology, the enzyme transcribed from gene $G_2$ reconverts $M_3$ to $M_1$, but the regulation of $G_1$ and $G_2$ remains dependent on $M_2$.
For each of our new topological models of the metabolator, the corresponding GM was assembled (see Appendix C) and numerical simulations completed to sample the space of parameters in the model. In each instance of the simulation, a random sample from the entire possible space of $\theta$ parameters was drawn. In order to efficiently sample the numerical space, we sampled uniformly from the ranges  $\theta_{r2}=(0,2], \theta_{r1}=[-2,0)$ and $\theta_i = (0,1]$ for all other elasticities. Since the parameters $\frac{M_{1,0}}{v},\frac{M_{2,0}}{v},$ and $\frac{1}{\beta}$ remained unconstrained, we decided to sample these parameters across seven orders of magnitude (from $10^{-3}$ to $10^3$ 1/sec, assuming $M_{1,0} = M_{2,0}$). For each instance of our simulation, the eigenvalues of $J$ were calculated. The process was repeated 20,000 times for each choice of $\frac{M_{1,0}}{v},\frac{M_{2,0}}{v}$ and $\frac{1}{\beta}$. In total, over $10^6$ simulations were conducted. Putative Hopf bifurcations were detected by examining whether a pair of complex conjugate eigenvalues existed which were sufficiently close to the imaginary axis (in our case, the magnitude of their real component was required to be less than $10^{-3}$). The size of the dynamical system (greater than four state-space variables) made confirming these results using the Routh-Hurwitz criterion analytically intractable.
The results of our numerical studies were unexpected. We found that reintroducing $G_3$ had no bearing on the appearance of a Hopf bifurcation. In fact, the model remained entirely stable, failing to exhibit a single instance of an eigenvalue with positive real part. On the other hand, the refined three-metabolite model did exhibit instability and Hopf bifurcations. These results suggest that the precise nature of the regulation is crucial for the metabolator to exhibit oscillations. In order for a Hopf bifurcation to be possible, the regulatory metabolite could not be repressing the expression of its own degrading enzyme. This delayed positive feedback appears to be a crucial element endowing the metabolator with the potential to undergo a Hopf bifurcation.

\section{Reverse Metabolator}
\label{sec:RM}
The difficulty in constructing synthetic circuits with poorly studied components often motivates synthetic biologists to construct several alternative designs. Frequently, these alternative designs are simply composed of rewired versions of the same regulatory components. In light of this, we investigated the dynamical behavior of one such re-wired version of the core metabolator, which we dub the reverse metabolator (RM). In the RM, the regulatory connections found in the core metabolator between $M_2$ and the two genes are reversed. Thus, $M_2$ now activates $G_1$, and hence its own production, and represses $G_2$, and hence its own degradation (see Figure 1D). We selected this design precisely because synthetic biologists frequently build a number of alternative constructs in the course of circuit assembly, in order to increase the odds of identifying a succesful design.

In nondimensional form, the RM is identical to (\ref{eq:met_linear}) but with the regulation of $g_1$ and $g_2$ reversed:

\begin{subequations}
	\begin{equation}
	\frac{dg_1}{dt} = \frac{m_2^2}{1+m_2^2} - g_1
	\label{eq:revmetl_g1}
	\end{equation}
	\begin{equation}
	\frac{dg_2}{dt} = \frac{1}{1+m_2^2} - g_2
	\label{eq:revmetl_g2}
	\end{equation}
	
	\begin{equation}
	\frac{dc}{dt} = \rho(1-m_2) 
	\label{eq:revmet1_c}
	\end{equation}
	\begin{equation}
	\frac{dm_2}{dt} = -\rho m_2 + \eta_1g_1(c-m_2) - \eta_2 g_2m_2.
	\label{eq:revmet1_n}
	\end{equation}
\label{eq:revmet}
\end{subequations}

\subsection{Hopf Bifurcations in the Reverse Metabolator}
For the RM, we repeated the explicit dynamical analysis outlined in Section 2. The fixed points of (\ref{eq:revmet}) are identical to those of the core metabolator ($g_1^* = g_2^* = \frac{1}{2}, m_2^* = 1$ and $c^* = \frac{\eta_1+\eta_2+2\rho}{\eta_1}$), but the Jacobian becomes

$$ \left[ \begin{array}{cccc}
-1 & 0 & 0 & \frac{1}{2} \\
0 & -1 & 0 & \frac{-1}{2} \\
0 & 0 & 0 & -\rho \\
2\rho+\eta_2 & -\eta_2 & \frac{\eta_1}{2} & \frac{-2\rho-\eta_1-\eta_2}{2} \end{array} \right].
$$

The characteristic equation of the RM is now

\begin{align}
(\lambda+1) \left(\lambda^3 + \lambda^2 \left( \frac{2+2\rho+\eta_1+\eta_2}{2} \right) + \lambda \left( \frac{\eta_1-\eta_2+\eta_1\rho}{2} \right) + \frac{\eta_1\rho}{2} \right) = 0.
\label{eq:revmet_char}
\end{align}

Next, we calculate the necessary condition for (\ref{eq:revmet_char}) to have a pair of purely imaginary roots (i.e. that $c_0 = \omega^2c_2 \text{ and } c_1 = \omega^2 c_3$, where $c_i$ is the coefficient of the $\lambda^i$th term of the cubic polynomial in (\ref{eq:revmet_char})). Assuming that $\lambda_{1,2} = \pm i\omega$, we find two conditions:

\begin{align}
\omega^2 = \frac{\eta_1\rho}{2+2\rho+\eta_1+\eta_2}, \omega^2 = \frac{\eta_1(1+\rho)-\eta_2}{2}.
\end{align}

Equating these two expressions for $\omega^2$, we arrive at a quadratic polynomial in $\eta_1$ 

\begin{align}
\eta_1^2(1+\rho) + \eta_1 \left( \rho(\eta_2+2\rho+2)+2 \right) -\eta_2(\eta_2+2\rho+2) = 0.
\label{eq:cond_revmet}
\end{align}

Using Descartes' rule of signs, (\ref{eq:cond_revmet}) must contain positive, real roots since $\eta_2>0$ and the $\eta_1^0$ term is negative. Therefore, there exists an open set of points $(\rho^*, \eta_1^*,\eta_2^*)$ at which the reverse metabolator undergoes a Hopf bifurcation. Using the numerical continuation software package AUTO \cite{auto}, a number of Hopf bifurcations spanning several orders of magnitude in $\eta_1, \eta_2$, and $\rho$ can be identified. 

\begin{figure}[ht]
	\centering
		\includegraphics[width=0.8\columnwidth]{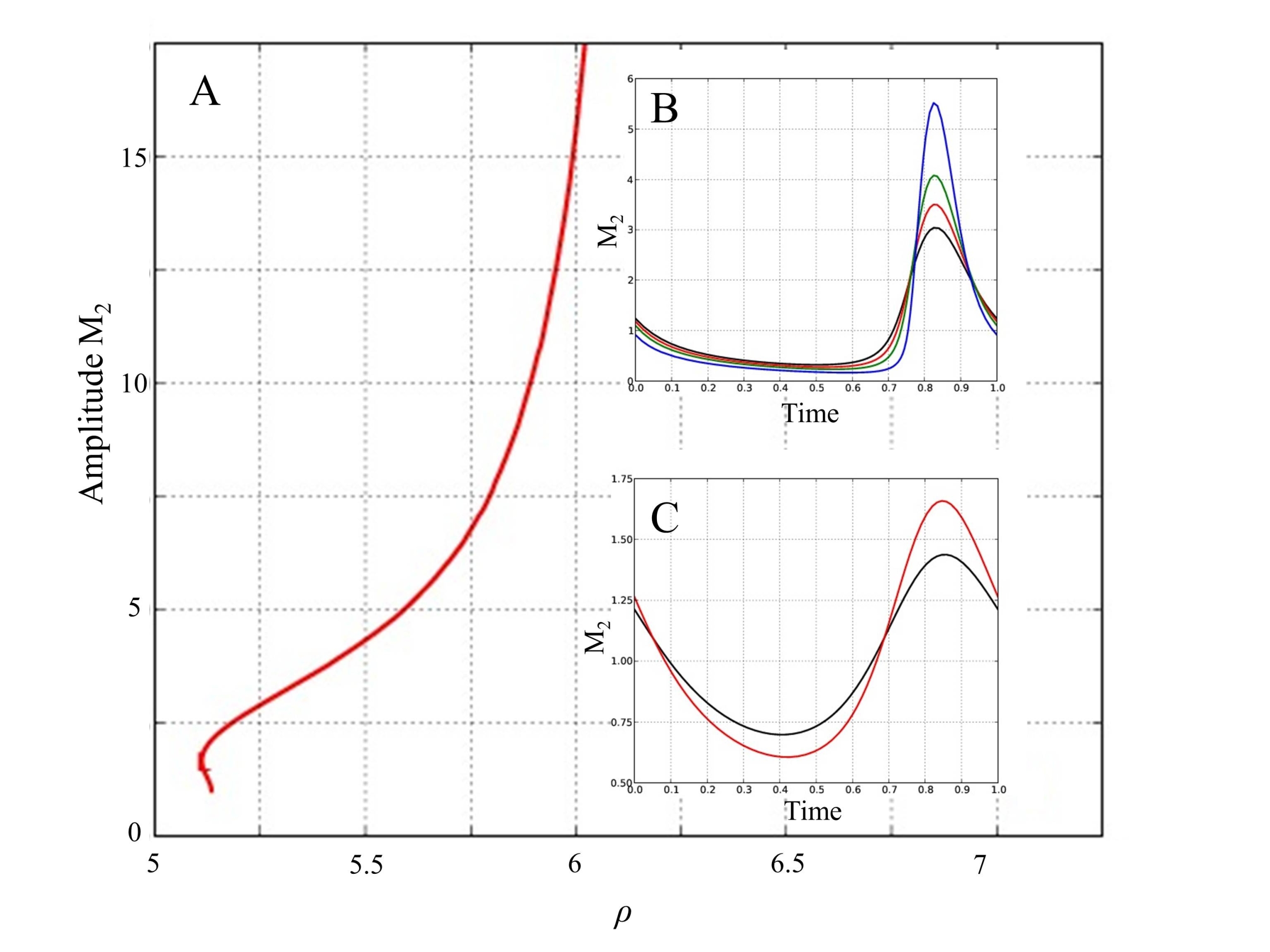}
	\label{fig:timeseries}
	\caption{(A) Continuation of a limit cycle from a Hopf bifurcation in the RM (\ref{eq:revmet}). For small values of $\rho$, two limit cycles exist (a large amplitude unstable cycle, and a small amplitude stable cycle), while for large values of $\rho$, a single, unstable limit cycle exists. (B) For $\rho$ = 5.5 (black), 5.6 (red), 5.8 (green), and 6.0 (blue), the unstable limit cycle exhibits increasingly sudden changes in abundance of $M_2$. (C) For $\rho = 5.15$, the two limit cycles (stable, black; unstable, red) exhibit smooth changes in amplitude. For all cases, $\eta_1 = 1.27, \eta_2 = 7.92$. }
\end{figure}

We used AUTO to study the behavior of the RM's limit cycles in different regions of parameter space. As shown in Figure 2, the limit cycles exhibit two notable features depending on the magnitude of inflow into the RM, characterized by the parameter $\rho$. First, for small $\rho$, two distinct periodic orbits exist: a small amplitude, stable limit cycle and a large amplitude unstable limit cycle. These two cycles annihilate at a fold of limit cycles. Both limit cycles display relatively smooth changes in amplitude (Figure 2C). Second, for large values of $\rho$, unstable periodic solutions showed characteristically sudden changes in amplitude, reminiscent of relaxation oscillations commonly found in neuroscience \cite{Izhikevich2006}. 

\subsection{Criticality of the Hopf Bifurcation}

\begin{figure}[ht!]
	\centering
		\includegraphics[width=0.8\columnwidth]{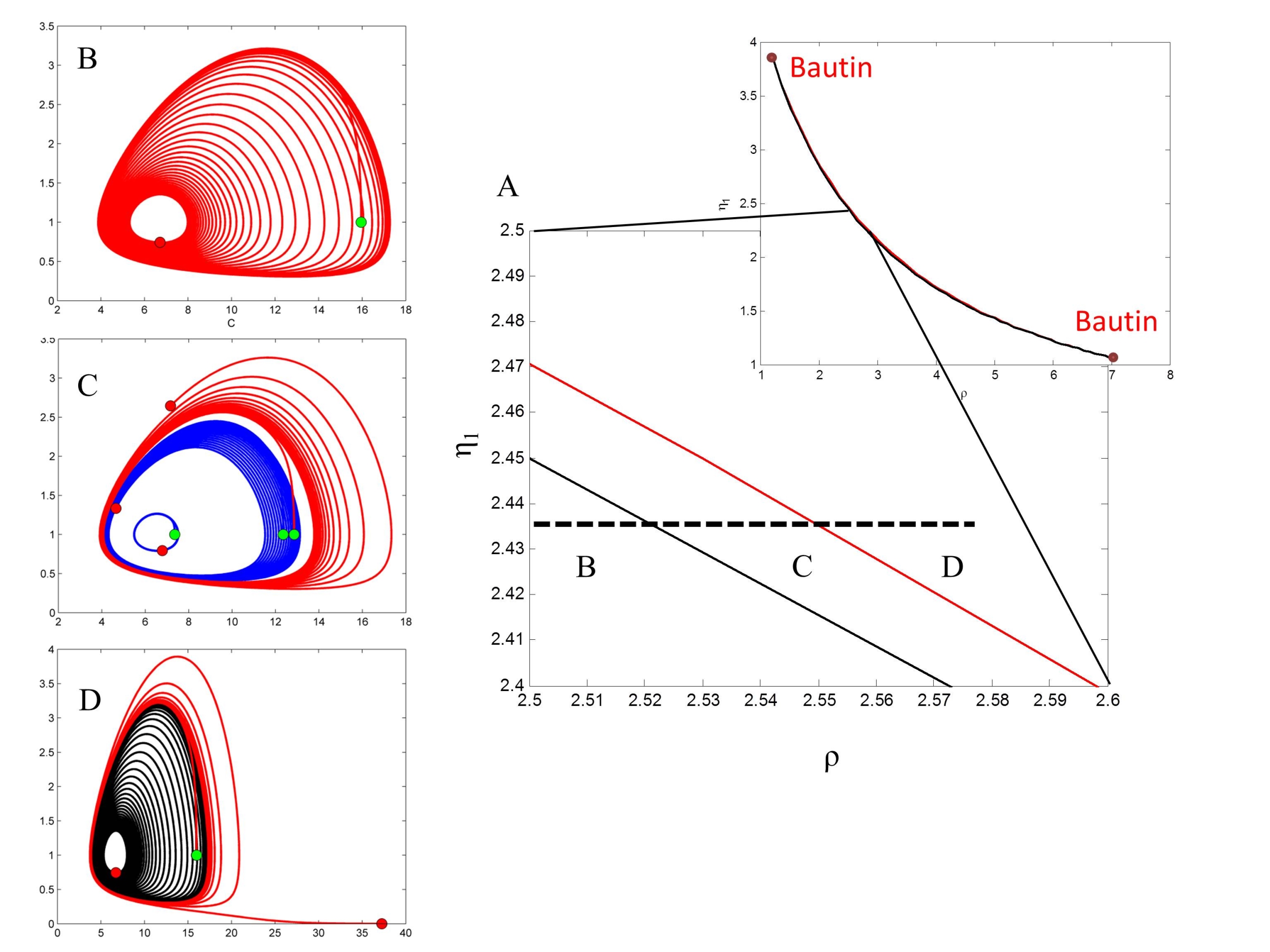}
	\label{fig:bif}
	\caption{(A) Two-parameter bifurcation diagram of the RM (\ref{eq:revmet}). As $\rho$ and $\eta_1$ are varied, a curve of Hopf bifurcations (red) and a fold of limit cycles (black) collide at a Bautin point. Insets (B), (C), and (D) correspond to the dynamics of the RM when $\eta_1 = 2.435$, in between the two Bautin points. Phase portraits correspond to the dynamics of metabolite $M_2$ (vertical axis) and total metabolite $C = M_1 + M_2$ (horizontal axis), with green dots indicating initial conditions and red dots indicating final positions. All simulations were run for a time interval of 1000. Black curves indicate trajectories approaching a fixed point, blue curves correspond to trajectories approaching a stable limit cycle, and red curves correspond to trajectories spiralling away from an unstable limit cycle.(B) For small $\rho$, the RM exhibits dampled, spiralling oscillations to a stable fixed point. (C) For $\rho = 2.55$, two distinct limit cycles exist. (D) For large values of $\rho$, the stable limit cycle in (C) undergoes a Hopf bifurcation and creates a stable fixed point, surrounded by a large amplitude unstable limit cycle. }
\end{figure}

We investigate the stability of the limit cycles generated by the Hopf bifurcations in the RM. Limit cycles arising from supercritical Hopf bifurcations are stable to small perturbations, while those generated by subcritical Hopf bifurcations are not. To explore the criticality of the observed Hopf bifurcations, a two-parameter bifurcation analysis of the RM was conducted in AUTO, shown in Figure 3A. As the two parameters $\rho$ and $\eta_1$ were varied, the criticality of the Hopf bifurcation was tracked using the magnitude of the leading Floquet multiplier. We observed that as the value of $\rho$ was decreased, the Hopf bifurcation switched from being subcritical to being supercritical. As the value of $\rho$ was decreased further, a second transition in criticality was observed, and the Hopf bifurcations returned to being subcritical. 

The point at which a Hopf bifurcation switches criticality is called a Bautin bifurcation. This higher-order bifurcation is associated with the collision of a fold of limit cycles with a manifold of Hopf bifurcations.  Using AUTO, we were able to follow the fold of limit cycles as it branched off from one Bautin bifurcation, tracked closely with the manifold of Hopf bifurcations, and reintersected with the Hopf curve at the second Bautin point (Figure 3A). The fold of limit cycles is the boundary between two regions of parameter space where distinct qualitative dynamical behaviors of occur. On one side of the fold (in this case, for Region B in Figure 3), no sustained oscillations exist. On the other side of the fold (Region C in Figure 3), a pair of limit cycles appears. For the RM, this pair contained a small amplitude stable limit cycle and a large amplitude unstable limit cycle. 

The existence of a Bautin bifurcation indicates there are two distinct routes to oscillation in the reverse metabolator. First, by decreasing $\rho$ for a fixed $\eta_1,\eta_2$ (moving from point D to point C in Figure 3A), the reverse metabolator passes through a supercritical Hopf bifurcation. In this case, oscillations appear at infinitesimal amplitude, and their amplitude grows as $\rho$ is decreased further. This type of bifurcation is identical to the one observed in the \textit{in vivo} metabolator \cite{metab}.  A second, and potentially more interesting, route to oscillation appears if the reverse metabolator passes through a fold of limit cycles. As the value of  $\rho$ is increased for fixed $\eta_1,\eta_2$ (moving from point B to C in Figure 3A), a finite-amplitude stable limit cycle appears, surrounded by a larger amplitude unstable limit cycle. Points perturbed from the unstable steady state will immediately be attracted to oscillations that are much larger in magnitude than those observed in the immediate vicinity of a Hopf bifurcation. Perturbations outside of the basin of attraction of the stable limit cycle will lead to spiraling instability, and eventually to the extinction of one of the metabolites.
We examine the biological implications of alternative modes of oscillation in the RM in the Discussion, Section 7.

\section{Generalized Model of the Reverse Metabolator}
The appearance of relaxation oscillations inspired us to investigate the role time scales play in the dynamics of the RM. The existence of such divergent time scales is well-known in the regulation of metabolism. The allosteric feedback of small metabolites and the post-translational modification of enzymes target the activity of an enzyme pool at time-scales on the order of microseconds \cite{Dill2011, Alon2006, Papin2005, Picard2009}. In contrast, the transcriptional regulation of enzyme levels (which occurs through the binding of protein transcription factors to DNA sequences) takes place at a substantially slower rate, on the order of minutes \cite{Alon2006, Schwanhausser2011,Heinemann2010}, and affects the actual size of the enzyme pool directly. Remarkably, the rate of an enzymatically catalyzed reaction itself, which the above processes ultimately regulate, is known to take place at characteristic rates spanning seven orders of magnitude, from microseconds to minutes \cite{Bar-Even2011}. 

In order to study the inherent time scales of the RM, we constructed a generalized model. This model was identical in structure to (\ref{eq:SKM_core2}), except that the signs of the elements corresponding to the transcriptional regulation of $M_2$ were reversed ($\theta_{r1} > 0, \theta_{r2}<0$). Our intent was to use the entries in the Jacobian $J$ to identify time-scales which characterized the dynamics of metabolites and genes in the RM. To understand the connection between $J$ and time-scales, recall that each row in $J$ describes the linearized dynamics of a small perturbation away from the equilibrium of one state variable. For example, the first row of $J$ corresponds to the dynamics of a perturbation in $M_1$:

\begin{equation}
\frac{d(\delta m_1)}{dt} = \frac{-\theta_1(1+\alpha)v}{M_{1,0}}\delta m_{1,0} + \frac{\theta_2v}{M_{1,0}}\delta m_2 + \frac{-(1+\alpha)v}{M_{1,0}}\delta g_1 + \frac{v}{M_{1,0}} \delta g_2 .
\label{eq:deltaM1}
\end{equation}
where $\delta m_1, \delta m_2, \delta g_1$, and $\delta g_2$ correspond to perturbations in $m_1, m_2, g_1$, and $g_2$, respectively. Upon inspection of (\ref{eq:deltaM1}), all terms on the right hand side share a common divisor $\frac{M_{1,0}}{v}$ which has units of time. We can treat this shared prefactor as a natural measure of the time scale of $M_1$. Repeating this process for all of the rows in (\ref{eq:J_core}), we identified the characteristic time scale for each state space variable. Entries in rows one and two of $J$, corresponding to the dynamics of the two metabolites, shared common divisors $\frac{M_{1,0}}{v}$ and $\frac{M_{2,0}}{v}$, respectively. Similarly, entries in the final two rows of $J$, corresponding to the genetic elements of the metabolator, shared common divisors $\frac{1}{\beta_1}$ and $\frac{1}{\beta_2}$ (again with units of time). 

\begin{figure}[ht!]
	\centering
		\includegraphics[width=\columnwidth]{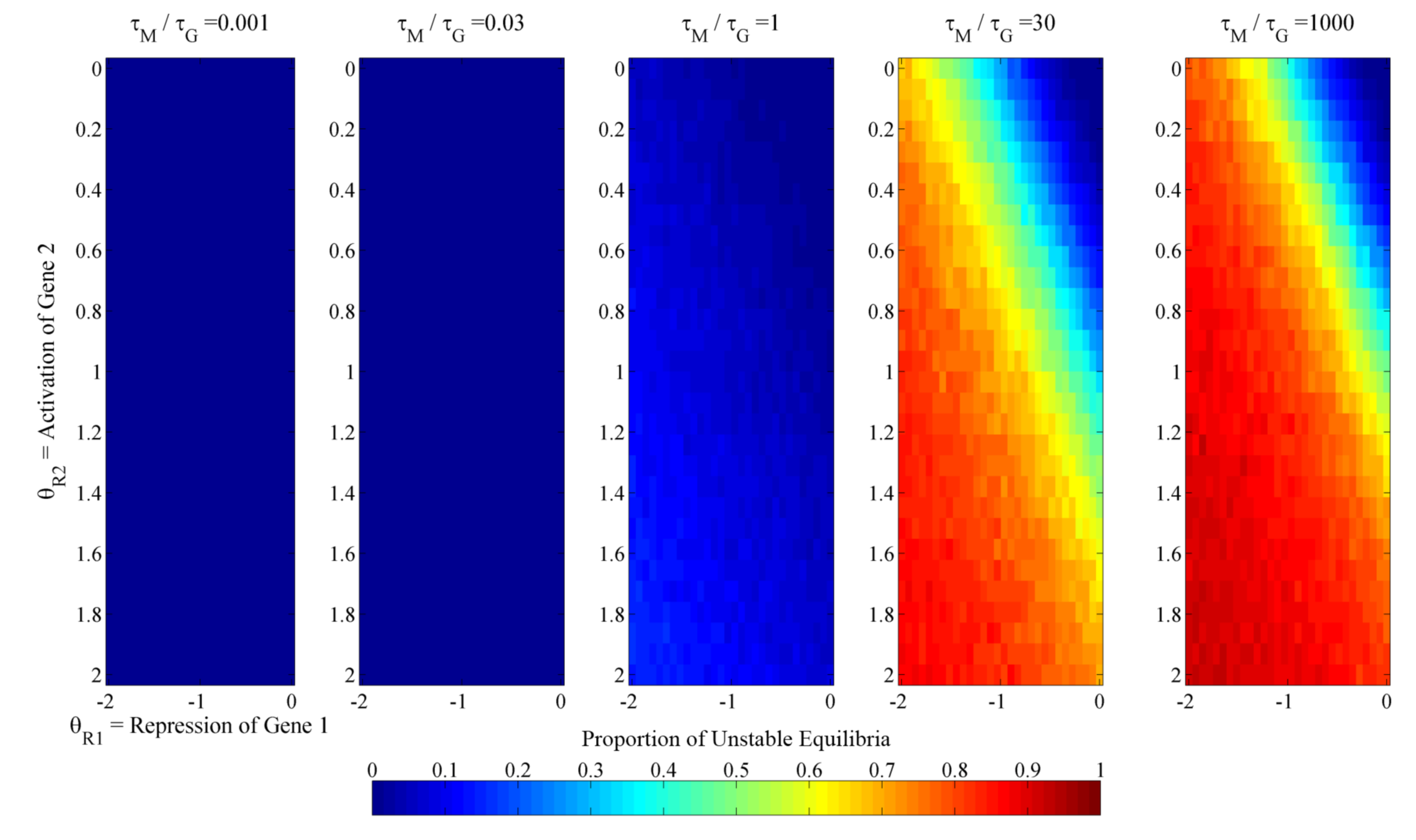}
	\label{fig:timescales}
	\caption{Dependence of stability of the RM on metabolic and genetic time scales (red = lower stability). Instability in the reverse metabolator increases as the genetic time scale becomes faster. For simplicity, we assumed that $M_1 = M_2, \beta_1=\beta_2$, and $v=1$, so that there existed only two time scales, one metabolic and one genetic. Additional numerical investigations indicate results are identical when this assumption is relaxed (results not shown).}
\end{figure}

We studied how the stability of $J$ depends on the ratio of the metabolic and genetic time scales. To do so, we generated a large number of random instances of $J$ and calculated the real component of the leading eigenvalue $\lambda_{max}$ of $J$ (a positive value indicated instability, while a negative value indicated stability). The sampling was completed by selecting a particular choice of the ratio of metabolic and genetic time scales, and then creating a 30x30 grid of $\theta_{r1}$ and $\theta_{r2}$ values. For each point on the grid, we sampled 1000 instances of all other $\theta$'s uniformly from the interval $ [0,1]$ for each point in the grid. The results of this simulation are shown in Figure 4. 

Two major trends regarding the stabiilty of the RM emerged from our simulations. First, we found that as the dynamics of the genetic elements in the reverse metabolator grew slower, the stability of the circuit as a whole increased. As the value of $\frac{1}{\beta}$ grew smaller and eventually became faster than the metabolic time scale, the stability of the circuit was observed to fall significantly. Second, our simulations also revealed that small regulatory elasticities contributed to the stability of the circuit, while large elasticities significantly reduced the probability of a stable circuit. This observation qualitatively agrees with our first observation relating high values of $\beta$ to instability. Large magnitude regulatory elasticities increase the same terms in the Jacobian as $\beta$. Together, our observations suggest that exceptionally slow regulation in the metabolator stabilizes the circuit, a conclusion that parallels commonly held notions on the speed of regulation and the role it plays in stability \cite{Klipp2009}.

\section{Conclusions and Discussion}

Contemporary biology is charged with understanding how fundamentally different cellular processes such as signalling, metabolism, and transcriptional regulation, interact and function as a coherent whole. Our work here explored the different types of conclusions which may be drawn from detailed kinetic models and more generic generalized models in the study of a hybrid metabolic-genetic oscillator. We first showed that the core metabolator (\ref{eq:met_linear}) has a unique fixed point which is stable for all parameter values (recall the end of Section 2). Hence, the core metabolator cannot undergo Hopf bifurcation to oscillations. Then, we developed a GM model of the core metabolator (\ref{eq:SKM_core1}), and we showed that this system also cannot undergo Hopf bifurcations by applying the Routh-Hurwitz criterion to the Jacobian (\ref{eq:J_core}) (see Section 3.2 and Appendix B). 

These analytical results for the core metabolator (\ref{eq:met_linear}) and its generalized model (\ref{eq:SKM_core1}) suggested that the network topology and the number of components in the network play an important role in determining whether or not oscillations are possible. As a result, in Section 4, we explored two more detailed versions of the core metabolator. The first of these has an intermediary delay gene $G_3$ (Figure 1B), but did not exhibit Hopf bifurcations in any of the large number of numerical simulations with randomly chosen system parameters we carried out. The second of these consists of an extra metabolite, which is also present in the synthetically constructed circuit \cite{metab}, effectively ``un-lumping'' $M_2$. This enhanced network exhibited Hopf bifurcations and instability for many of the parameter values we simulated. The delayed positive feedback created by the presence of this third metabolite is responsible for the induced oscillations. Moreover, this result also suggests that the self-repression of $M_2$ induced by the network topologies for the core metabolator, generalized model of the core metabolator, and the first ``enhanced'' network (Figures 1A and 1B), is responsible for the absence of oscillations in those systems. 

Theory and simulation play important roles in synthetic biology. As we have shown, they can highlight generic designs which have certain desirable behaviors, \textit{e.g.} oscillations, and they can help exclude those that do not. In this respect, the analysis of the enhanced metabolator model with a third metabolite (Figure 1C) suggested that we explore some other four-component models with related designs that do exhibit oscillations. In particular, we studied system (\ref{eq:revmet}), the reverse metabolator, in which the wiring from $M_2$ to the two genes is reversed compared to the core metabolator, as shown in Figure 1D. This fundamentally altered the system dynamics so that $M_2$ now activates, rather than represses, the gene $G_1$ and hence $M_2$ is now self-activating. The analysis of Section 5 of the RM reveals that it exhibits both local and global routes to oscillations, via Hopf bifurcations and saddle-node bifurcations of limit cycles, respectively, as shown in Figure 3.

Most of the experimentally observed oscillations in synthetic circuits have arisen through local Hopf bifurcations \cite{Elowitz2000, metab, Purcell2010}. This is very likely to be due to the comparable ease of predicting local bifurcations. Indeed, the GM methods which are used throughout our work are only capable of identifying global bifurcations in very special and exceedingly rare cases (\textit{i.e.} when they are known to occur near the intersection of two local bifurcations, like a double Hopf bifurcation). In contrast to the Hopf (where oscillations appear at infinitesimal amplitude), oscillations arising from a fold of limit cycles (as seen in the RM) immediately appear at a finite, non-zero amplitude. This feature may make such a ``global'' route to oscillation desirable in synthetic biology, which is still tackling the problem of designing robust cellular oscillators. The similarity between the RM and the original metabolator (ultimately, simply just a switch of promoters in the architecture of the circuit) suggests that constructing the RM and investigating its dynamics \textit{in vivo} may be relatively easy, and may lead to the experimental observation of a global bifurcation.

Finally, we propose that the construction of large generalized models of biological circuits may enable the resolution of several open problems regarding the dynamics of large, complex systems. We investigated one such question here (recall Section 6): the role that time scales play in determining the stability of the system. While we found that stability generally increased as the time scale of regulation became longer (Figure 4), it remains unclear how general this conclusion is among biological networks. A closely related question is whether strongly stabilizing components exist within biological networks. Such components have been argued for in ecological networks \cite{Gross2009}, as well as in prior work on metabolism using GM \cite{SKM}. However, the extent to which components of distinct types of biological networks (\textit{e.g.} signalling, transcriptional regulation, post-translational modifications) act to stabilize or destabilize metabolism is unknown. The investigation of such questions, aided by generalized modeling, seems potentially tractable numerically; computational toolboxes for the automated assembly and inspection of generalized models already exist \cite{Girbig2012}, and their extension to GM with regulation appears straightforward. 

\begin{acknowledgments}
ER and DS were supported by grants from the Office of Science (BER), U.S. Department of Energy (DE-SC0004962) and National Science Foundation NSF DMS-0602204 EMSW21-RTG Biodynamics at Boston University. TJK was supported in part by NSF grant DMS-1109587. The funders had no role in study design, data collection and analysis, decision to publish, or preparation of the manuscript. ER would like to acknowledge the kind support of the Santa Fe Institute while studying the metabolator at the 2011 Summer School.
\end{acknowledgments}

\bibliographystyle{plain}

\clearpage
\appendix

\section{Appendix: Formulation and Non-dimensionalization of the Core Metabolator}
We begin by writing out the equations for the metabolator. In dimensionalized form, they are
\begin{subequations}
	\begin{equation}
	\frac{dG_1}{dt} = \frac{\alpha_1}{K_1^2 + M_2^2} - \beta_1G_1 \equiv F_1
	\label{eq:dim_g1}
	\end{equation}
	\begin{equation}
	\frac{dG_2}{dt} = \frac{\alpha_2M_2^2}{K_2^2 + M_2^2} - \beta_2G_2 \equiv F_2
	\label{eq:dim_g2}
	\end{equation}
	\begin{equation}
	\frac{dM_1}{dt} = V_{in} - v_1G_1M_1 + v_2G_2M_2 \equiv F_3
	\label{eq:dim_M}
	\end{equation}
	\begin{equation}
	\frac{dM_2}{dt} = -V_{out}M_2 + v_1G_1M_1 - v_2G_2M_2 \equiv F_4.
	\label{eq:dim_N}
	\end{equation}
\label{eq:mdim}
\end{subequations}

Let's begin nondimensionalizing $F_1$. Let $m_2 = \frac{M_2}{K_1}$. We can rewrite (\ref{eq:dim_g1}) as 

\begin{align}
\frac{1}{\beta_1}\frac{dG_1}{dt} &= \frac{\alpha_1}{ \beta_1K_1^2(1+m^2) } - G_1 \\
\frac{1}{K_1\beta_1}\frac{dG_1}{dt} &= \frac{\alpha_1}{\beta_1K_1^3(1+m_2^2)} - \frac{G_1}{K_1} \\
\frac{dg_1}{d\tau} &= \frac{\gamma_1}{1 + m_2^2} - g_1,
\label{eq:g1_nd}
\end{align} 

\noindent where $\gamma_1 = \frac{\alpha_1}{\beta_1K_1^3}, g_1 = \frac{G_1}{K_1}$ and $\tau = \beta_1t$.

Then, the nondimensionalized equation for $G_2$ immediately follows as
\begin{equation}
c_1\frac{dg_2}{d\tau} = \frac{\gamma_2c_2^2m_2^2}{1+c_2^2m_2^2} - g_2,
\label{eq:g2_nd}
\end{equation}

\noindent where $g_2 = \frac{G_2}{K_1}, c_1 = \frac{\beta_1}{\beta_2}, c_2 = \frac{K_1}{K_2}$ and $\gamma_2 = \frac{\alpha_2}{\beta_2K_1}$.

For (\ref{eq:dim_M}), we rewrite it by substituting those nondimensionalized variables and parameters which we have already defined and choose to scale $M_1$ so that $m_1 = \frac{M_1}{K_1}$. Doing so, we get
\begin{align}
\beta_1K_1 \frac{dm_1}{d\tau} &= V_{in} - v_1g_1m_1K_1^2 + v_2g_2m_2K_1^2 \\
\frac{dm_1}{d\tau} &= \rho_1 - \eta_1g_1m_1 + \eta_2g_2m_2,
\label{eq:m_nd}
\end{align}

\noindent where $\rho_1 = \frac{V_{in}}{\beta_1K_1}, \eta_1 = \frac{v_1K_1}{\beta_1}$ and $\eta_2 = \frac{v_2K_1}{\beta_1}$.

Finally, the nondimensionalized equation for $M_2$ follows simply and is
\begin{equation}
\frac{dm_2}{d\tau} = -\rho_2m_2 + \eta_1g_1m_1 - \eta_2g_2m_2,
\label{eq:n_nd}
\end{equation}

\noindent where $\rho_2 = \frac{V_{out}}{\beta_1}$.

We also note that a simpler way to define the system is in terms of ``total'' metabolite in the system $C = M_1 + M_2$. Then, $\dot{C} = \dot{M_1} + \dot{M_2} = V_{in} - V_{out}M_2$. We can non-dimensionalize this system as well to find that 
\begin{equation}
\frac{dc}{d\tau} = \rho_1 - \rho_2m_2,
\label{eq:c_nd}
\end{equation}

\noindent where $c = \frac{C}{K_1}$.

For the purpose of simplifying the analysis in the article, we let $c_1=c_2=\gamma_1 = \gamma_2 = 1$ and $\rho_1 = \rho_2=\rho$. Then, the system formed by (\ref{eq:g1_nd}),(\ref{eq:g2_nd}),(\ref{eq:c_nd}), and (\ref{eq:n_nd}) is the nondimensionalized core metabolator we study in this article.
\section{Appendix: Proof that Generalized Model of the Core Metabolator Does Not Admit Hopf Bifurcations}
We work with the Jacobian of the generalized model of the core metabolator (\ref{eq:J_core}):

\begin{equation}
J-\lambda I = \left[ \begin{array}{cccc}
\frac{-\theta_1(1+\alpha)v}{M_{1,0}}- \lambda & \frac{\theta_2v}{M_{1,0}} & \frac{-(1+\alpha)v}{M_{1,0}} & \frac{v}{M_{1,0}} \\
\frac{\theta_1(1+\alpha)v}{M_{2,0}} & \frac{-\theta_2v-\alpha\theta_3v}{M_{2,0}}- \lambda & \frac{(1+\alpha)v}{M_{2,0}} & \frac{-v}{M_{2,0}} \\
0 & \beta_1\theta_{r1} & -\beta_1- \lambda & 0 \\
0 & \beta_2\theta_{r2} & 0 & -\beta_2- \lambda \end{array} \right],
\label{eq:J_corenew}
\end{equation}

\noindent where we have let $\beta_i = \frac{L_i\theta_{d,i}}{G_{i,0}}$ and $\theta_{r,i} = \frac{\theta_{f,i}}{\theta_{d,i}}$. This means that $\theta_{r1} < 0, \theta_{r2}>0$ and removes two otherwise free parameters from the analysis.

Adding multiples of one row to another does not change the value of the determinant. Thus, to simplify matters, we multiply row one of (\ref{eq:J_corenew}) by $\frac{M_{1,0}}{M_{2,0}}$ and add it to row 2 of (\ref{eq:J_corenew}) to find

\begin{equation}
J-\lambda I = \left[ \begin{array}{cccc}
\frac{-\theta_1(1+\alpha)v }{M_{1,0}}- \lambda & \frac{\theta_2v}{M_{1,0}} & \frac{-(1+\alpha)v}{M_{1,0}} & \frac{v}{M_{1,0}} \\
\frac{-\lambda M_{1,0}}{M_{2,0}} & \frac{-\alpha\theta_3v}{M_{2,0}}- \lambda & 0 & 0 \\
0 & \beta_1\theta_{r1} & -\beta_1- \lambda & 0 \\
0 & \beta_2\theta_{r2} & 0 & -\beta_2- \lambda \end{array} \right].
\label{eq:J_corenew2}
\end{equation}

Now expanding the determinant of (\ref{eq:J_corenew2}) along the first column, we find

\begin{align}
|J-\lambda I| =& \left(\frac{-\theta_1(1+\alpha) v}{M_{1,0}} - \lambda \right)\left( \frac{-\alpha \theta_3 v}{M_{2,0}}-\lambda \right)(-\beta_1 - \lambda )(-\beta_2 - \lambda) + \nonumber \\
& \frac{M_{1,0}}{M_{2,0}}\lambda \Big| \begin{array}{ccc}
\frac{\theta_2v}{M_{1,0}} & \frac{-(1+\alpha)v}{M_{1,0}} & \frac{v}{M_{1,0}} \\
\beta_1\theta_{r1} & -\beta_1- \lambda & 0 \\
\beta_2\theta_{r2} & 0 & -\beta_2- \lambda \end{array} \Big|.
\end{align}

Expanding again, we find
\begin{align}
|J-\lambda I| = & \left(\frac{-\theta_1(1+\alpha) v}{M_{1,0}} - \lambda \right)\left( \frac{-\alpha \theta_3 v}{M_{2,0}}-\lambda \right)(-\beta_1 - \lambda )(-\beta_2 - \lambda) + \nonumber \\
& \frac{M_{1,0}}{M_{2,0}}\lambda \left( \frac{v}{M_{1,0}}(\beta_2\theta_{r2})(\beta_1+\lambda) + (-\beta_2-\lambda) \left( \frac{\theta_2 v}{M_{1,0}}(-\beta_1-\lambda) + \frac{\beta_1\theta_{r1}(1+\alpha)v}{M_{1,0}}  \right)  \right).
\end{align}

Expanding this expression a final time and grouping terms, we find the coefficients of the characteristic equation for the Jacobian $J$ of the core metabolator are (where $a_i$ corresponds to the coefficient of the $\lambda^{i}$th term):
\begin{align}
	a_4 &= 1 \nonumber \\
	a_3 &= \beta_1 + \beta_2 + \frac{(1+\alpha)v\theta_1}{M_{1,0}} + \frac{v\theta_2 + \alpha v\theta_3}{M_{2,0}} \nonumber \\
	a_2 &= \beta_1\beta_2 + \frac{\alpha(1+\alpha)v^2\theta_1\theta_3}{M_{1,0}M_{2,0}} + (\beta_1+\beta_2) \left( \frac{(1+\alpha)v\theta_1}{M_{1,0}} + \frac{v\theta_2+\alpha v\theta_3}{M_{2,0}} \right) + \frac{\beta_2v}{M_{2,0}}(\theta_{r2}) + \frac{\beta_1v}{M_{2,0}} (- (1+\alpha)\theta_{r1} ) \nonumber \\
	a_1 &= \beta_1\beta_2v \left(\frac{(1+\alpha)\theta_1}{M_{1,0}} + \frac{\alpha\theta_3 + \theta_{r2} + \theta_2 - (1+\alpha)\theta_{r1}}{M_{2,0}} \right) + (\beta_1+\beta_2)\frac{\alpha(1+\alpha)v^2\theta_1\theta_3}{M_{1,0}M_{2,0}} \nonumber \\
	a_0 &= \frac{\alpha(1+\alpha)\beta_1\beta_2v^2\theta_1\theta_3}{M_{1,0}M_{2,0}}.
\end{align}

The Routh Hurwitz criterion states that all of the eigenvalues of a 4x4 matrix have negative real parts when the following conditions on its characteristic polynomial are satisfied
\begin{align}
a_n > 0, n = 0,1,2,3,4 \nonumber \\
a_2a_3 > a_1a_4 \nonumber \\
a_1a_2a_3 > a_1^2a_4 + a_0a_3^2.
\end{align}

 First, note that the condition $a_n>0$ is satisfied. Next, we calculate the quantity $a_2a_3$:

\begin{align}
\label{eq:a2a3}
a_2a_3 = (\beta_1	\beta_2) & \left(\beta_1 + \beta_2 + \frac{(1+\alpha)v\theta_1}{M_{1,0}} + \frac{v\theta_2 + \alpha v\theta_3}{M_{2,0}} \right) \\
+ \frac{\alpha(1+\alpha)v^2\theta_1\theta_3}{M_{1,0}M_{2,0}} & \left(\beta_1 + \beta_2 + \frac{(1+\alpha)v\theta_1}{M_{1,0}} + \frac{v\theta_2 + \alpha v\theta_3}{M_{2,0}} \right) \nonumber \\ 
+ (\beta_1 + \beta_2) \left(\frac{(1+\alpha)v\theta_1}{M_{1,0}} + \frac{v\theta_2 + \alpha v\theta_3}{M_{2,0}} \right) & \left(\beta_1 + \beta_2 + \frac{(1+\alpha)v\theta_1}{M_{1,0}} + \frac{v\theta_2 + \alpha v\theta_3}{M_{2,0}} \right) \nonumber \\
+(\frac{\beta_2v\theta_{r2}}{M_{2,0}} - \frac{(1+\alpha)v\beta_1\theta_{r1}}{M_{2,0}} ) & \left(\beta_1 + \beta_2 + \frac{(1+\alpha)v\theta_1}{M_{1,0}} + \frac{v\theta_2 + \alpha v\theta_3}{M_{2,0}} \right) \nonumber.
\end{align}

Now, $a_1a_4$ = $a_1$ because $a_4 = 1$. Explicitly calculating $a_2a_3 - a_4$:
\begin{align}
\label{eq:a2a3minus}
a_2a_3 - a_1a_4 = &\beta_1\beta_2(\beta_1+\beta_2) \\
+&\frac{\alpha(1+\alpha)v^2\theta_1\theta_3}{M_{1,0}M_{2,0}} \left(\frac{(1+\alpha)v\theta_1}{M_{1,0}} + \frac{v\theta_2 + \alpha v\theta_3}{M_{2,0}} \right) \nonumber \\
+&(\beta_1 + \beta_2) \left(\frac{(1+\alpha)v\theta_1}{M_{1,0}} + \frac{v\theta_2 + \alpha v\theta_3}{M_{2,0}} \right) \left(\beta_1 + \beta_2 + \frac{(1+\alpha)v\theta_1}{M_{1,0}} + \frac{v\theta_2 + \alpha v\theta_3}{M_{2,0}} \right) \nonumber \\
+&(\frac{\beta_2v\theta_{r2}}{M_{2,0}} - \frac{(1+\alpha)v\beta_1\theta_{r1}}{M_{2,0}} ) \left( \frac{(1+\alpha)v\theta_1}{M_{1,0}} + \frac{v\theta_2 + \alpha v\theta_3}{M_{2,0}} \right) +\frac{\beta_2^2v\theta_{r2}}{M_{2,0}} - \frac{(1+\alpha)v\beta_1^2\theta_{r1}}{M_{1,0}} \nonumber .
\end{align}

All the terms are positive, so the condition $a_2a_3 > a_1a_4$ is satisfied. 

Now, we rewrite the final condition as $a_1(a_2a_3 - a_1a_4) - a_0a_3^2 > 0$. Calculating $a_0a_3^2$, we find:

\begin{align}
\label{eq:a0a3}
a_0a_3^2 =& \frac{(\beta_1+\beta_2)^2\alpha(1+\alpha)\beta_1\beta_2v^2\theta_1\theta_3}{M_{1,0}M_{2,0}} \\
& + 2(\beta_1+\beta_2) \left( \frac{(1+\alpha)v\theta_1}{M_{1,0}} + \frac{v\theta_2 + \alpha v\theta_3}{M_{2,0}} \right) \left( \frac{ \alpha(1+\alpha)\beta_1\beta_2v^2\theta_1\theta_3}{M_{1,0}M_{2,0}} \right) \nonumber \\
& + \left( \frac{(1+\alpha)v\theta_1}{M_{1,0}} + \frac{v\theta_2 + \alpha v\theta_3}{M_{2,0}} \right)^2 \left( \frac{ \alpha(1+\alpha)\beta_1\beta_2v^2\theta_1\theta_3}{M_{1,0}M_{2,0}} \right) \nonumber .
\end{align}

Also, rewriting $a_1$ and highlighting the two key components
\begin{align}
a_1 = \overbrace{ \beta_1\beta_2v \left(\frac{(1+\alpha)\theta_1}{M_{1,0}} + \frac{\alpha\theta_3 + \theta_{r2} + \theta_2 - (1+\alpha)\theta_{r1}}{M_{2,0}} \right) }^\text{Term 1} + \overbrace{ (\beta_1+\beta_2)\frac{\alpha(1+\alpha)v^2\theta_1\theta_3}{M_{1,0}M_{2,0}} }^\text{Term 2}.
\label{eq:a1}
\end{align}

Now, note that the first line of (\ref{eq:a2a3minus}) multipled by Term 2 of (\ref{eq:a1}) gives the first term of (\ref{eq:a0a3}). Similarly, we can isolate a term like $2(\beta_1+\beta_2)\left( \frac{ \alpha(1+\alpha)v^2\theta_1\theta_3}{M_{1,0}M_{2,0}} \right)$ from the third line of (\ref{eq:a2a3minus}) and multiply it by Term 1 of (\ref{eq:a1}) to obtain the second term of (\ref{eq:a0a3}), so that all the remaining terms of the product of the two lines are positive. Finally, we can multiply the second line of (\ref{eq:a2a3minus}) by Term 1 of (\ref{eq:a1}) to give the third term of (\ref{eq:a0a3}) and have all leftover terms be positive. The remaining terms are all positive, proving that the Routh-Hurwitz criteria are satisfied and all of the eigenvalues of the Jacobian have negative real part.

\section{Appendix: Generalized Models of Other Metabolators}
For the metabolator in Figure 1C, we used the GM

\begin{align}
	\frac{dM_1}{dt} &= I + G_2R_2(M_2) - G_1R_1(M_1) \nonumber \\
	\frac{dM_2}{dt} &= G_1R_1(M_1) - G_2R_2(M_2) - R_3(M_2) \nonumber \\
	\frac{dG_1}{dt} &= P_1(-G_3) - D_1(G_1) \nonumber \\
	\frac{dG_2}{dt} &= P_2(M_2) - D_2(G_2) \nonumber \\
	\frac{dG_3}{dt} &= P_3(M_2) - D_3(G_3).
\label{eq:SKM_aug1}
\end{align}

The Jacobian of (\ref{eq:SKM_aug1}) is

\begin{equation}
J = \left[ \begin{array}{ccccc}
\frac{-\theta_1(1+\alpha)v }{M_{1,0}} & \frac{\theta_2v}{M_{1,0}} & \frac{-(1+\alpha)v}{M_{1,0}} & \frac{v}{M_{1,0}} & 0 \\
\frac{\theta_1(1+\alpha)v}{M_{2,0}} & \frac{-\theta_2v-\alpha\theta_3v}{M_{2,0}} & \frac{(1+\alpha)v}{M_{2,0}} & \frac{-v}{M_{2,0}} & 0\\
0 & 0 & -\beta_1 & 0 & \beta_1\theta_{r1}\\
0 & \beta_2\theta_{r2} & 0 &-\beta_2 & 0 \\
0 & \beta_3\theta_{r3} & 0 & 0 & -\beta_3 \\ \end{array} \right].
\label{eq:J_aug1}
\end{equation}

We assume again that $\theta_{r1}<0$ and all other elasticities are greater than zero.

For the metabolator in Figure 1D, we use the GM:

\begin{align}
	\frac{dM_1}{dt} &= I + G_2R_2(M_3) - G_1R_1(M_1) \nonumber \\
	\frac{dM_2}{dt} &= G_1R_1(M_1) - R_4(M_2) \nonumber \\
	\frac{dM_3}{dt} &= R_4(M_2) - G_2R_2(M_3) - R_3(M_3) \nonumber \\
	\frac{dG_1}{dt} &= P_1(-M_2) - D_1(G_1) \nonumber \\
	\frac{dG_2}{dt} &= P_2(M_2) - D_2(G_2).
\label{eq:SKM_aug3}
\end{align}

The Jacobian of (\ref{eq:SKM_aug3}) is

\begin{equation}
J = \left[ \begin{array}{ccccc}
\frac{-\theta_1(1+\alpha)v }{M_{1,0}} & 0 & \frac{\theta_2v}{M_{1,0}}  & \frac{-(1+\alpha)v}{M_{1,0}} & \frac{v}{M_{1,0}} \\
\frac{\theta_1(1+\alpha)v}{M_{2,0}} & \frac{-\theta_4(1+\alpha)v}{M_{2,0}} & 0 & \frac{(1+\alpha)v}{M_{2,0}} & 0 \\
0 & \frac{\theta_4(1+\alpha)v}{M_{3,0}} & \frac{-\theta_2v-\alpha\theta_3v}{M_{3,0}} & 0 & \frac{-v}{M_{3,0}} \\
0 & \beta_1\theta_{r1} & 0 & -\beta_1 & 0 \\
0 & \beta_2\theta_{r2} & 0 & 0 & -\beta_2 \end{array} \right].
\label{eq:J_aug2}
\end{equation}

Above, $\theta_4$ corresponds to the elasticity of reaction $R_4(M_2)$. We again assume again that $\theta_{r1}<0$ and all other elasticities are greater than zero.

\end{document}